\documentclass{nature}

\def\specchar#1{{\sc #1}}

\def\SiI{\mbox{Si\,\specchar{i}}}

\def\HeI{\mbox{He\,\specchar{i}}}
\usepackage{lineno}
\usepackage{graphicx}
\makeatletter
\let\saved@includegraphics\includegraphics
\AtBeginDocument{\let\includegraphics\saved@includegraphics}
\renewenvironment*{figure}{\@float{figure}}{\end@float}
\makeatother

\title{Signatures of sunspot oscillations and the case for chromospheric resonances}

\author{T. Felipe$^{1,2}$}

\begin{document}

\maketitle

\begin{affiliations}
 \item Instituto de Astrof\'{\i}sica de Canarias, 38205, C/ V\'{\i}a L{\'a}ctea, s/n, La Laguna, Tenerife, Spain
 \item Departamento de Astrof\'{\i}sica, Universidad de La Laguna, 38205, La Laguna, Tenerife, Spain
\end{affiliations}

\begin{abstract}

Sunspots host a large variety of oscillatory phenomena, whose properties depend on the nature of the wave modes and the magnetic and thermodynamic structure of the spot. Umbral chromospheric oscillations exhibit significant differences compared to their photospheric counterparts. They show an enhanced power and a shorter dominant period, from waves with an amplitude of a few hundred meters per second in the five-minute band at the photosphere, to amplitudes of several kilometers per second in the three-minute band at the chromosphere. Various models have been proposed to explain this behaviour\cite{Fleck+Schmitz1991, Centeno+etal2006, Chae+etal2019}, including the presence of a chromospheric resonance cavity between the photosphere and the transition region\cite{Zhugzhda+Locans1981,Zhugzhda2008,Botha+etal2011, Snow+etal2015, Felipe2019}. Jess et al.\cite{Jess+etal2019} claimed the detection of observational evidence supporting this model, obtained from the comparison of spectropolarimetric observations and numerical simulations. Here, it is shown that the observational insight reported by Jess et al. is not a common property of sunspots. More importantly, numerical modelling also shows that it is not an unequivocal signature of an acoustic resonator.    

\end{abstract}

\begin{figure}
\includegraphics[width=16cm]{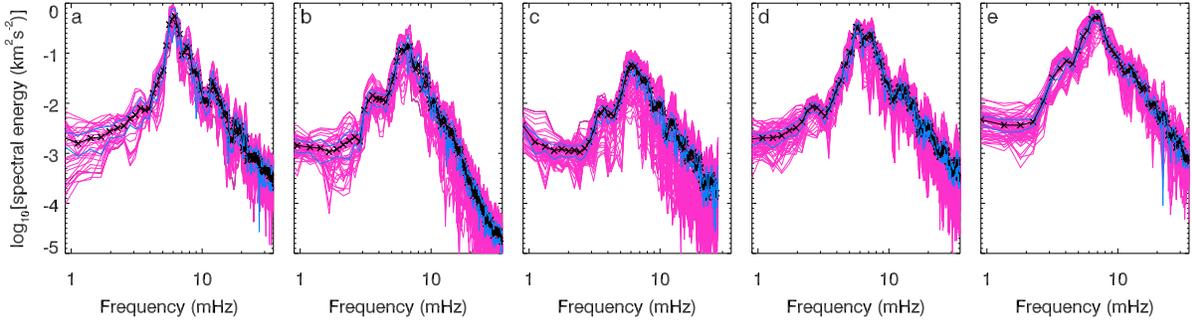} % this command will be ignored
\caption{{\bf Umbral chromospheric spectral energies of five observed time series}. Each panel shows the \HeI\ 10830 \AA\ spectral energy of a different sunspot displayed on log-log axes. The pink lines correspond to the spectra calculated from an individual point inside the umbra. The black lines with crosses represent the average umbral spectral energy. The blue lines show the 99\% confidence interval. Panels {\bf a-c} illustrate data acquired with TIP/VTT from sunspots NOAA 09173 (2000 Oct 1, a)\cite{Centeno+etal2006}, NOAA 09443 (2001 May 9, {\bf b})\cite{Centeno+etal2006}, and NOAA 10969 (2007 Aug 28, {\bf c})\cite{Felipe+etal2010b}. Panels {\bf d} and {\bf e} show data obtained with GRIS/GREGOR from the sunspot NOAA 12662 on 2017 Jun 17 ({\bf d}) and 2017 Jun 18 ({\bf e})\cite{Felipe+etal2018b}.}
\label{fig:obs}
\end{figure}

Jess et al. analysed a temporal series of the chromospheric \HeI\ 10830 \AA\ triplet, whose formation height is around 2,100 km above the solar surface\cite{Avrett+etal1994}. From the analysis of the umbral Doppler velocity, they found a power enhancement at 20 mHz. The presence of this prominent peak is used as the sole indicator of a resonant cavity. Figure \ref{fig:obs} illustrates the power spectra of the umbral velocity measured from the \HeI\ 10830 \AA\ triplet in five independent sunspot observations. These data were acquired with either the Tenerife Infrared Polarimeter at the Vacuum Tower Telescope (TIP/VTT) or the GREGOR Infrared Spectrograph attached to the GREGOR telescope (GRIS/GREGOR), both telescopes located at the Spanish Observatorio del Teide, Tenerife. They have already been analysed in previous works\cite{Centeno+etal2006, Felipe+etal2010b, Felipe+etal2018b}. All the umbral chromospheric power spectra show a similar trend, with a peak at the three-minute band and a gentle power reduction for higher frequencies. None of them exhibits the peak at 20 mHz reported by Jess et al. This result is in agreement with previous chromospheric sunspot observations, which do not present any hint of a power excess around 20 mHz\cite{Lites1986}. This power enhancement is a specific feature found in a single sunspot observation, rather than a standard property that could be potentially employed for seismological analyses of solar active regions.

Regarding the modelling, previous works have not found any indication of high-frequency power strengthening as a direct result of the presence of chromospheric resonances in the sunspot atmosphere. Theoretical developments of the spectrum produced by waves partially trapped due to the temperature gradients predict the presence of several peaks at specific frequencies. Their power is reduced with frequency as a result of the rapid decrease with frequency of the incident wave power from subphotospheric layers\cite{Zhugzhda2008}. Several numerical simulations of chromospheric resonances above sunspot umbrae also show no trace of a power increase at high frequencies\cite{Botha+etal2011, Felipe2019}. On the contrary, the paper by Jess et al. identifies this power increase as a signature of the chromospheric resonance. This claim is based on the comparison of two simulations with and without an upper temperature gradient. Since only the simulation with the transition region exhibits a power peak at around 20 mHz, the presence of this power in their observations is taken as a proof of the resonance. 

\begin{figure}
\includegraphics[width=13cm]{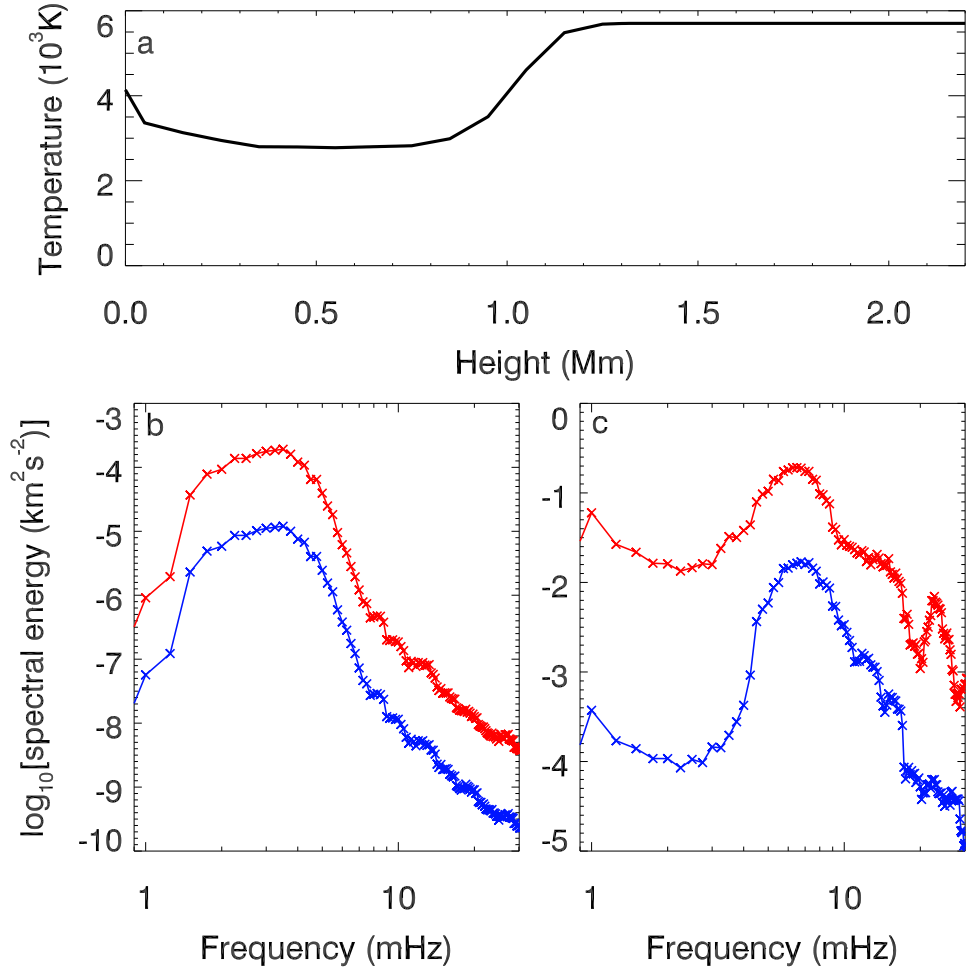} % this command will be ignored
\caption{{\bf Temperature stratification and spectral energies of numerical simulations of wave propagation through an atmosphere without upper temperature gradient}. Panel {\bf a} shows the temperature stratification of the umbral atmospheric model from Avrett\cite{Avrett1981}, which has been rescaled to slightly reduce the temperature at the temperature minimum. Panels {\bf b} and {\bf c} illustrate the spectral energy at the photosphere ({\bf b}) and at the formation height of the \HeI\ 10830 \AA\ line ({\bf c}). In both panels, the red lines correspond to a simulation where the driver of the waves exhibits realistic amplitudes, whereas the simulation illustrated with the blue lines employed the same driver with a reduced amplitude in order to keep the simulations in the linear regime.}
\label{fig:sim}
\end{figure}

A numerical simulation of wave propagation in an umbral atmospheric model without upper temperature gradient has been performed. Similarly to Jess et al., it is driven by the photospheric velocity extracted from observations of the \SiI\ 10827 \AA\ line in a sunspot umbra\cite{Felipe+etal2018b}. Figure \ref{fig:sim}a illustrates the temperature stratification of the atmosphere. It corresponds to the model by Avrett \cite{Avrett1981}, whose temperature at the temperature minimum has been slightly reduced. The red lines from Fig. \ref{fig:sim}b,c show the spectral energy of the vertical velocity at the photosphere and the chromosphere, respectively. The chromospheric spectrum exhibits a significant power enhancement at 22 mHz, similar to that detected by Jess et al. in their observations. The photospheric spectrum shows that this power excess is not present in the driver, but appears as a result of upward wave propagation from the photosphere to the chromosphere, without requiring the action of a steep temperature gradient at the transition region. This simulation proves that the presence in the chromospheric spectra of a power peak at around 20 mHz cannot be interpreted as a signature of a chromospheric resonance cavity. 

Another simulation has been performed using the same background model and driver but reducing the amplitude of the latter in order to maintain the simulations in the linear regime. The power spectra of this numerical experiment are plotted in blue in Fig. \ref{fig:sim}b,c. During this simulation no shocks are formed, opposite to the previous case. The absence of the high-frequency power excess in the chromospheric spectrum suggests that this feature is the outcome of non-linearities. The power spectrum of this linear simulation is in qualitative agreement with the simulation without transition region presented by Jess et al. However, discrepancies are found when waves with realistic amplitudes (velocity oscillations up to 10 km/s at the chromosphere) are computed. This lack of agreement can be due to the use of different background atmospheres and wave drivers, or it could be related to differences in the numerical tools employed for the computations.

In the simulations without transition region, the frequency of the main chromospheric power peak (around 5-8 mHz) is given by the power distribution of waves travelling from deeper layers (introduced by the driver) and the stratification of the atmosphere. The temperature minimum is especially relevant since it determines the maximum value of the cutoff frequency. Only waves with frequency above the cutoff frequency can freely propagate to upper atmospheric layers and become dominant at the chromosphere. As they develop into shocks, power enhancements are expected to appear at certain frequencies, corresponding to the harmonics of the main oscillatory frequencies\cite{Chae+etal2018}. The chromospheric high-frequency spectrum is then determined by an interplay between those harmonics and the power from the high-frequency waves which directly propagate from deeper layers. If a steep temperature gradient is present at the transition region, the power at some specific frequencies is enhanced as a result of the resonances\cite{Felipe2019} and, thus, the power of their harmonics might stand out above the signal of the propagating high-frequency waves. The chromospheric resonance can favour the appearance of those high-frequency peaks, but it is not a mandatory requirement.

Jess et al.\cite{Jess+etal2019} presented an extraordinary data set of wave propagation in sunspot umbrae. They reported for the first time a striking chromospheric power enhancement at around 20 mHz. This singular data set can potentially provide new insights about sunspot wave propagation, although it cannot be employed as a reference to describe the general properties of sunspot oscillations due to its uniqueness (see Fig. \ref{fig:obs} for a sample of chromospheric power spectra). Therefore, their interpretation of the power peak at around 20 mHz is too speculative. The numerical modelling developed in this work proves that their evidence is not robust enough to claim it as a proof of resonances in a chromospheric cavity since other processes can also generate that power excess. The transition region is a well-established property of the solar atmosphere, and its steep temperature gradient must certainly play a role in the observed chromospheric waves. More analyses are required to understand its effects on sunspot oscillations.

%% Put the bibliography here, most people will use BiBTeX in
%% which case the environment below should be replaced with
%% the \bibliography{} command.

\begin{addendum}
\item[Data availability] 
The data that support this paper are available from the corresponding author upon reasonable request.

\end{addendum}

\bibliographystyle{naturemag}
\bibliography{biblio.bib}

% \begin{thebibliography}{1}
% \bibitem{dummy} Articles are restricted to 50 references, Letters
% to 30.
% \bibitem{dummyb} No compound references -- only one source per
% reference.
% \end{thebibliography}

%\bibliography{sample}

%% Here is the endmatter stuff: Supplementary Info, etc.
%% Use \item's to separate, default label is "Acknowledgements"
-
\begin{addendum}
 \item Financial support from the State Research Agency (AEI) of the Spanish Ministry of Science, Innovation and Universities (MCIU) and the European Regional Development Fund (FEDER) under grant with reference PGC2018-097611-A-I00 is gratefully acknowledged. The 1.5-meter GREGOR solar telescope was built by a German consortium under the leadership of the Leibniz-Institute for Solar Physics (KIS) in Freiburg with the Leibniz Institute for Astrophysics Potsdam, the Institute for Astrophysics Göttingen, and the Max Planck Institute for Solar System Research in Göttingen as partners, and with contributions by the Instituto de Astrofísica de Canarias and the Astronomical Institute of the Academy of Sciences of the Czech Republic. The author wishes to acknowledge the contribution of Teide High-Performance Computing facilities to the results of this research. TeideHPC facilities are provided by the Instituto Tecnol\'ogico y de \hbox{Energ\'ias} Renovables (ITER, SA). URL: http://teidehpc.iter.es. 

 \item[Competing Interests] The author declares that he has no competing financial interests.
 \item[Correspondence] Correspondence and requests for materials
should be addressed to T. Felipe.~(email: tobias@iac.es).
\end{addendum}

%%
%% TABLES
%%
%% If there are any tables, put them here.
%%

\end{document}